\documentclass[floatfix,twocolumn,aps,prd,showpacs,amsmath,amssymb,nofootinbib]{revtex4}

\usepackage[utf8]{inputenc}
\usepackage{epsfig}
\usepackage{float}
\usepackage{slashed}

\begin{document}

\title{
{Effects of the Pseudo-Chern-Simons action for strongly correlated electrons in the plane}
}

    \author{R. F. Ozela} \email{ozelarf@gmail.com}
\affiliation{Faculdade de F\'{\i}sica, Universidade Federal do Par\'a,
  66075-110 Bel\'em, PA,  Brazil} 

\author{Van S\'ergio Alves} \email{vansergi@ufpa.br}
\affiliation{Faculdade de F\'{\i}sica, Universidade Federal do Par\'a,
  66075-110 Bel\'em, PA,  Brazil}

\author{G. C. Magalh\~aes} \email{gabrielconduru@hotmail.com}
\affiliation{Faculdade de F\'{\i}sica, Universidade Federal do Par\'a,
  66075-110 Bel\'em, PA,  Brazil}

\author{Leandro O. Nascimento} \email{lon@ufpa.br}
\affiliation{Faculdade de Ci\^encias Naturais, Universidade Federal do Par\'a,
  68800-000 Breves, PA,  Brazil}


\date{\today}

\begin{abstract}

Chiral symmetry breaking comes from the mass dynamically generated through interaction of Dirac fermions for both quantum electrodynamics in (2+1)D (QED3) and (3+1)D (QED4). In QED3, the presence of a Chern-Simons (CS) parameter affects the critical structure of the theory, favoring the symmetric phase where the electron remains massless.
 Here, we calculate the main effects of a Pseudo-Chern-Simons (PCS) parameter $\theta$ into the dynamical mass generation of Pseudo quantum electrodynamics (PQED). The $\theta$-parameter provides a mass scale for PQED at classical level and appears as the pole of the gauge-field propagator. After calculating the full electron propagator with the Schwinger-Dyson equation at quenched-rainbow and large-$N$ approximations, we conclude that $\theta$ affects the critical parameters related to the fine-structure constant, $\alpha_c(\theta)$, and to the number of copies of the matter field, $N_c(\theta)$, by favoring the symmetric phase. In the continuum limit ($\Lambda\to \infty$), nevertheless, the $\theta$-parameter do not affect the critical parameters. We also compare our analytical results with numerical findings of the integral equation for the mass function of the electron.
 
\end{abstract}



\maketitle

\section{Introduction} 

In the last decades, quantum field theories in (2+1)D have been extensively studied. This interest is partly due  to its potential applications in condensed matter physics \cite{girvin1987quantum,wilczek1990fractional,halperin1982quantized,
laughlin1981quantized,laughlin1988superconducting,chen1989anyon,farakos1998dynamical,herbut2002qed,franz2002qed,thouless1982quantized,haldane1988model,lee2006doping,affleck1988large,ioffe1989gapless,kim1997massless,kim1999theory,rantner2001electron,franz2001algebraic,herbut2002antiferromagnetism,liu2002chiral,liu2003effect,Bernevig,shun2018topological,hasan2010colloquium} and, for comparison with quantum chromodynamics, at low-energy scales. In particular, quantum electrodynamics in 2+1 dimensions (QED3) has waged interesting features that are similar to quantum chromodynamics, such as dynamical mass generation~\cite{pisarski1984chiral,appelquist1986spontaneous,appelquist1988critical,nash1989higher,hoshino1989dynamical,dagotto1989computer,dagotto1990chiral,pennington1991masses,curtis1992dynamical,atkinson1990dynamical,kondo1992cutoff,burden1991light} and confinement~\cite{burden1992photon,grignani1996confinement,maris1995confinement}.

PQED is the dimensional reduction of QED4, when the matter field is constrained to move within the plane and the photons are allowed to propagate away from and back to this plane; as such, it is a unitary~\cite{unitarity}  model that respects causality~\cite{do1992canonical,barci1996canonical} and describes mixed-dimensionality systems. Because of that, it has been applied to describe the electromagnetic interaction in 2D materials, such as graphene~\cite{grapexp,novoselov2005two,geim2010rise,review},  silicene~\cite{2Dmaterials}, and transition metal dichalcogenides~\cite{wang2018colloquium,2Dmaterials,li2014black,ye2014intrinsic,wang2012electronics}. Within the myriad of results it has given rise to, we allude in hindsight to dynamical mass generation for fermion at zero and finite temperature~\cite{CSBPQED,nascimento2015chiral,PhysRevD.96.034005}, interaction-driven quantum valley Hall effect~\cite{PRX2015}, quantum corrections of the electron g-factor~\cite{gfactor},  electron-hole pairing (excitons) in transition metal dichalcogenides~\cite{TMDPQED,chaves2017excitonic}, optical infrared conductivity of graphene~\cite{teber2017method}, emergence of a dynamically generated mass with Gross-Neveu interaction~\cite{fernandez2021dynamical}, Yukawa potential in the plane~\cite{Yukawa, pproca}, and PQED cavity effects~\cite{placa,cavity}.

For massless Dirac particles, the dynamical mass generation has been investigated in several scenarios in both QED4~\cite{johnson1964self,maskawa1974spontaneous,fukuda1976schwinger,curtis1990truncating,atkinson1993chiral,fomin1983dynamical,miransky1985dynamical,miransky1985dynamics,kogut1988new,kogut1989strongly,dagotto1991non,kondo1989phase,gusynin1990vacuum,bashir2011critical,kizilersu2009building} and QED3~\cite{pisarski1984chiral,appelquist1986spontaneous,appelquist1988critical,nash1989higher,hoshino1989dynamical,dagotto1989computer,dagotto1990chiral,pennington1991masses,curtis1992dynamical,atkinson1990dynamical,kondo1992cutoff,burden1991light}, providing a critical value either for the fine-structure constant or for the number of flavors, respectively. This non-perturbative effect is usually calculated with the Schwinger-Dyson equations for the full electron propagator~\cite{dyson1949s,schwinger1951green,roberts1994dyson}. For graphene,
this dynamical mass generation implies in a gap opening at the Dirac points of the quasiparticle excitation~\cite{PRX2015}, which might cause a topological phase transition~\cite{Bernevig}.

Meanwhile, at topological phase transition for the gauge field, the Chern-Simons (CS) term plays a vital role. Indeed, it generates a mass for the gauge field while breaking the parity symmetry and preserving the gauge symmetry~\cite{Dunne}. As a consequence, the CS term applications to the Meissner effect are vastly documented in the literature~\cite{carena1991superconductivity,fradkin1989jordan,
lykken1990anyonic,deser1982three,siegel1979unextended,jackiw1981super}.

The addition of the CS term to QED3 has been shown to favor the symmetric phase in both representations for spinors~\cite{hong1993dynamical,williams1994dyson,Kondo}, where the electron remains massless. 
On the other hand, coupled to PQED,  the CS term changes the electric permittivity of the vacuum -- effectively working as a dielectric medium~\cite{pqed+cs,carrington} -- and has been used to calculate the nonperturbative mass generation for the fermions~\cite{olivares2020influence}. This CS parameter, however, is dimensionless and can not generate a mass for the gauge field in PQED. In order to do so, one has to consider the Pseudo Chern-Simons (PCS) action,  obtained by dual transformation of an abelian CS-Higgs~\cite{PLBOzela} or by a bosonization  of massive Dirac fermions~\cite{GabrielPCS}. We shall refer as Pseudo Maxwell-Chern-Simons (PMCS) to the model that combines PQED and PCS; this PMCS has a peculiar feature of producing bound states of electrons~\cite{PLBOzela,GabrielPCS}.  Nevertheless, effects of dynamical mass generation had not been investigated for the PMCS until now.




In this work, we use the Schwinger-Dyson equations to investigate the dynamical mass generation associated with four-component fermions coupled to the PCS terms. We use the so-called quenched-rainbow approach and the large-$N$ expansion for calculating the mass function $\Sigma(p)$,\textit{ i.e.}, the term of the electron self-energy that yields the dynamical mass. Our main results show that there exist a critical coupling constant $\alpha_c(\theta)$ and a critical number of fermions  $N_c(\theta)$, where both of them separate the broken phase from the symmetric (massless) phase. It is shown that these are dependent on the PCS parameter $\theta$. Thereafter, we compare our approximated analytical results with the numerical results obtained from integral equation for the mass function and a very good agreement is found as well as a confirmation of the critical parameters. Because $\theta$ has dimension of mass, it essentially means an effective correlation length and, therefore, our results may be relevant for describing a symmetry restoration in two-dimensional materials due to a screening of the electromagnetic interaction. The precise microscopic mechanism that generates $\theta$ is obviously absent within our description, because we are assuming a continuum model,  unless by the presence of an ultraviolet (UV) cutoff.



This paper is organized as follow. In Sec. II we present the model and  we set up the coupled SD equations for the photon and fermion propagators . In Sec. III present the truncation scheme using the quenched-rainbow approximation and we  investigate the analytical solutions of the integral equation for the mass function of fermions.  In Sec. IV we examine the inﬂuence of vacuum polarization on the CSB using the N massless fermion ﬂavors version and adopting a $1/N$ expansion. We summarize our results in Sec. V.  In Appendix~\ref{appA}, we review some aspects of the criticality of QED3 coupled to the usual CS term.
 
\section{The model and its truncated Schwinger-Dyson equation}
\label{PCS-SDE}

We start with a CS term modified by a pseudo differential operator. This action has been shown dual to the Chern-Simons-Higgs model in Ref.~\cite{PLBOzela}), using standard path-integral formalism. When coupled to PQED, we find a Euclidean action, given by
\begin{equation}
    \begin{split}
        {\cal L}_{\mathrm{PMCS}}=\frac{1}{2} \frac{F_{\mu\nu}F^{\mu\nu}}{\sqrt{-\Box}}
+\frac{i\theta }{2} \frac{\epsilon_{\mu\nu\gamma}A^\mu \partial^\nu A^\gamma}{\sqrt{-\Box}}+
\\+\Bar{\psi}i \slashed{\partial}\psi 
+\frac{\lambda}{2} \frac{(\partial_\mu A^\mu)^2}{\sqrt{-\Box}}
+e\Bar{\psi}\gamma^\mu\psi A_\mu,
    \end{split}\label{LMCS}
\end{equation}  
 where
$A^\mu$ is a gauge field,
$\theta$ is the CS massive parameter, 
$e$ is the dimensionless coupling constant, 
$1/\sqrt{-\Box}$ is a pseudo-differential operator, and the fermions are represented by the 4-component spinors $\psi$ and $\Bar{\psi}$, describing the Dirac field. 
This gives rise to a bare gauge-field propagator given by
 \begin{equation}
 \Delta_{\mu\nu}^{(0)}(p)=\frac{p^2P_{\mu\nu}+\theta \epsilon_{\mu\nu\gamma}p^\gamma}{2\varepsilon\sqrt{p^2}(p^2+\theta^2)}+\Delta^{GF}_{\mu\nu}(p), \label{Photonprop}
 \end{equation}
 where $\varepsilon$ is included in order to describe the dieletric constant, and $\Delta^{GF}_{\mu\nu}(p)$ represents the gauge-dependent part which is null in Landau's gauge $\lambda \to\infty$. The fermion bare propagator is given by
 \begin{equation}
     S_F^{(0)}(p)=-\frac{\slashed{p}}{p^2},
 \end{equation}
and an interaction vertex by $\Gamma^{(0)}_\mu=e\gamma_\mu.$

 The main consequence of the $\theta$-parameter is to provide a bounded pair of electrons~\cite{PLBOzela}
within a typical distance of $r_0 \propto 1/\theta$ as explained in Ref.~\cite{GabrielPCS}. Furthermore,  such parameter does not change the main results of the renormalization group functions at one-loop expansion because it does appear in any
divergent term of the  two-point vertex functions associated with  Eq.~(\ref{LMCS}). Next, we shall calculate
the main contribution of $\theta$ through nonperturbative approach, provided by the Schwinger-Dyson equation in the ladder approximation.
\begin{figure}[htbp]
    \centering
    \noindent\includegraphics[width=.35\textwidth]{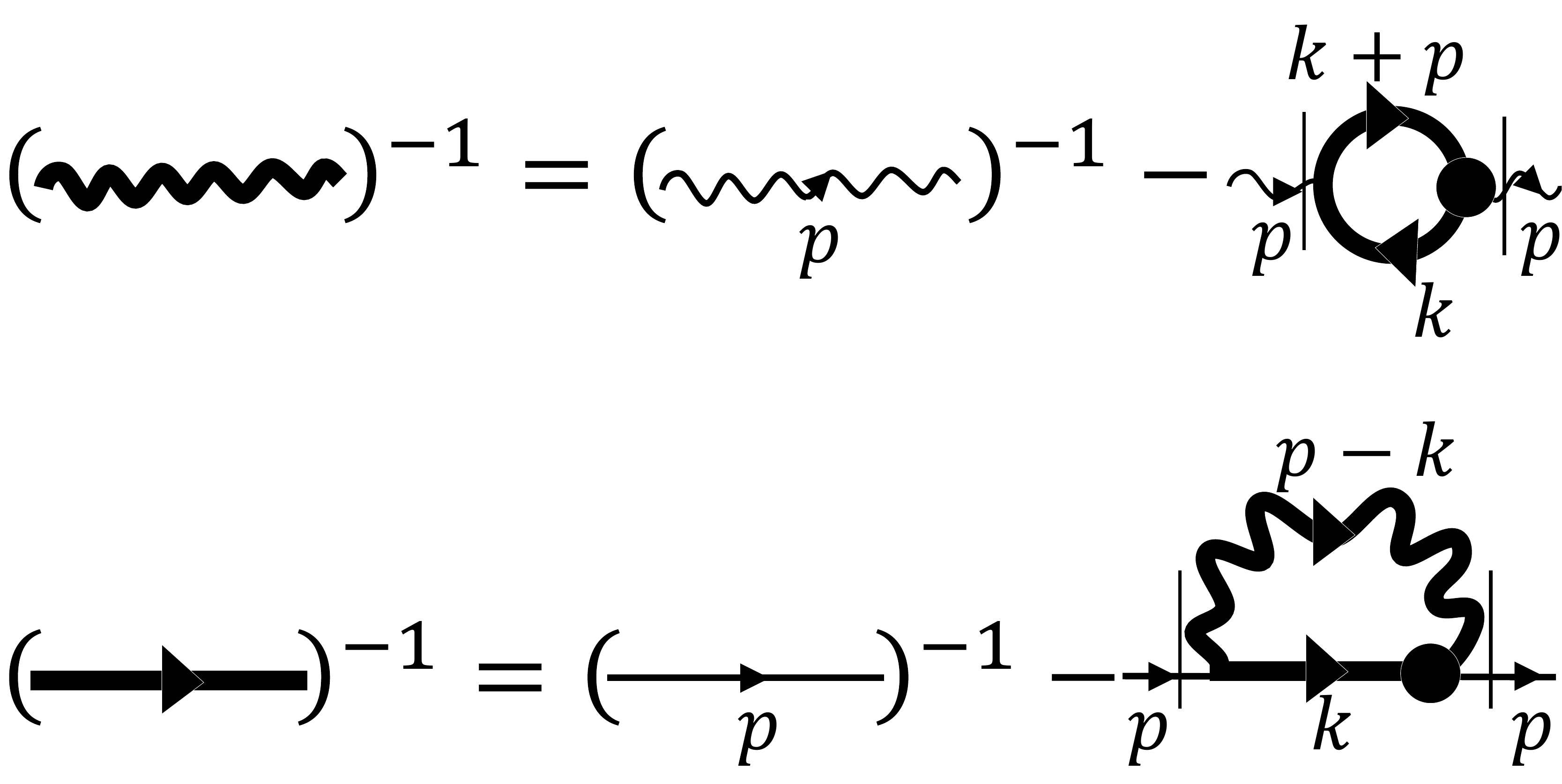}
  \caption{The fermion and photon SD equation, respectively. The bold and filled dots indicate full propagators and vertices. The second term on the righthand side, in both figures,  represents the photon self-energy $\Pi^{\mu\nu}(p)$ and the electron self-energy $\Xi(p)$.}
    \label{SDE}
\end{figure}

In order to investigate the corrections 
coming apace with the PMCS dynamics, we  consider the following coupled Schwinger-Dyson equations displayed in Fig.~\ref{SDE} and analytically  written as
\begin{equation}\label{selfelectron1}
    [S_F(p)]^{-1}=    [S_F^{(0)}(p)]^{-1}  -\Xi (p)
\end{equation}
and 
\begin{equation}
\Delta^{-1}_{\mu\nu}(p)=\Delta^{-1}_{0\mu\nu}(p) -\Pi_{\mu\nu}(p),    
\end{equation}\label{Sig}
where the electron and photon self-energies respectively reads
\begin{equation}\label{Sigma}    
    \Xi(p)=e^2\int\frac{d^3k}{(2\pi)^3}\gamma^\mu S_F(k) \Gamma^\nu(k,p)\Delta_{\mu\nu}(p-k),
\end{equation}
and
\begin{equation}\label{Pi}
\Pi^{\mu\nu}(p)= -e^2\int\frac{d^3k}{(2\pi)^3} {\rm Tr}\left[\gamma^{\mu} S_F(k+p)\Gamma^{\nu}(k,p)S_F(k)\right].   \end{equation}

In the Euclidean spacetime we can write, 
\begin{equation}\label{selfelectron2} [S_F(p)]^{-1}=-\gamma^\mu p_\mu A(p)+\Sigma(p),
\end{equation}
where $A(p)$ is the wave function renormalization and $\Sigma(p)$ is the mass function. Posteriorly, substituting Eq.(\ref{selfelectron2}) into Eq.(\ref{selfelectron1}), we find  that 
\begin{equation}
     \begin{split}
-\gamma^\alpha p_\alpha A(p)+\Sigma(p)&=-\gamma^\alpha p_\alpha+\\
- e^2\int\frac{d^3k}{(2\pi)^3}&\frac{\gamma^\mu S_F(k) \Gamma^\nu(k,p)\Delta_{\mu\nu}(p-k)}{\gamma^\beta p_\beta A(k)+\Sigma(k)}\label{gammamatrices}
    \end{split}
\end{equation}
From Eq.~(\ref{gammamatrices}), we can isolate  an expression for $A(p)$ by multiplying $\slashed{p}$ in both sides and then calculating the trace over the Dirac matrices, hence  \begin{equation}\label{Wavefunction}
 A(p)=1-\frac{e^2}{4p^2}
 \int\frac{d^3k}{(2\pi)^3}Tr\left[{\gamma^\beta p_\beta F(p)}\right],
 \end{equation}
 where for the sake of simplicity we define
\begin{equation}
F(p)=\left[\frac{\gamma^\mu S_F(k) \Gamma^\nu(k,p)\Delta_{\mu\nu}(p-k)}{\gamma^\beta p_\beta A(k)+\Sigma(k)}\right].
\end{equation}

For $\Sigma(p)$, we simply  take the trace over the $\gamma^\mu$ matrices in Eq.~(\ref{gammamatrices}) to get
 \begin{equation}
 \Sigma(p)
 =- e^2 \int\frac{d^3k}{(2\pi)^3}Tr[F(p)].
 \label{SigmaSemQuench}
\end{equation}

From now on, we shall consider a set of approximations in order to obtain an analytical result for the mass function. This allow us to discuss the dynamical breaking of chiral symmetry as well as to calculate the critical behavior of the model in terms of the $\theta$ parameter.

\section{Quenched-Rainbow Approximation}
\label{quenchedSec}
The ladder order is obtained by the so-called quenched-rainbow approximation. In the 4x4 representation it has been previously applied to investigate the criticality of the PQED ~\cite{CSBPQED,nascimento2015chiral,PhysRevD.96.034005}. It imposes that the full vertex $\Gamma^\mu$ and the full gauge-field propagator $\Delta_{\mu\nu}$ can be exchanged by their bare counterparts $\gamma^\mu$ and $\Delta_{\mu\nu}^{(0)}$. This approach decouples Eq.(\ref{Sigma}) and Eq.(\ref{Pi}). Also, note from Eq.(\ref{Wavefunction}) we can conclude that $A(p)\approx 1+{\cal O} (e^2)$. Keeping this in mind, and using the trace properties for the Dirac matrices, namely, $Tr\left(  \mathbb{1} \right)=4$, $Tr(\gamma^\mu\gamma^\nu)=-4\delta^{\mu\nu}$, and $Tr(\gamma^\mu\gamma^\nu\gamma^\beta)=0$, 
we may calculate $\Sigma(p)$ from Eq.(\ref{SigmaSemQuench}), hence,
\begin{equation}\label{Sigma3}
    {\Sigma(p)=e^2\int\frac{d^3k}{(2\pi)^3} \Sigma(k) \delta^{\mu\nu}\Delta^{(0)}_{\mu\nu}}(p-k).
    \end{equation}

Substituting Eq.(\ref{Photonprop}) into Eq.(\ref{Sigma3}) and using spherical coordinates, we obtain, in the Landau gauge 
\begin{equation}
\begin{split}
    \Sigma (p) = \frac{\alpha}{2\pi^2}  \int_0^\Lambda dk\,  \frac{ k^2 \Sigma (k) }{k^2+\Sigma^2 (k)} \int_0^{2\pi} d\phi\\ \int_0^\pi d\eta 
\frac{\sin(\eta)\sqrt{p^2+k^2- 2 p k \cos(\eta)}}{p^2+k^2- 2 p k \cos(\eta)+\theta^2},
\end{split}    
\end{equation}
where  $\Lambda$ is an UV cutoff, and $\alpha=e^2/4\pi\varepsilon$ is the fine-structure constant.
After solving the integrals over $\eta$ and $\phi$,
we find
\begin{equation}
\begin{split}
\Sigma (p) = \frac{\alpha}{\pi }   \int_0^\Lambda   \frac{k \Sigma (k) }{k^2+\Sigma^2 (k)}\Bigg\{ \frac{ \vert p + k \vert -\vert p - k \vert}{p }  +\\ + \frac{\theta}{p}\left[ \arctan\left(\frac{\vert p-k \vert}{\theta}\right)- \arctan\left(\frac{\vert k+p \vert}{\theta}\right)\right]\Bigg\}   dk.
\end{split}\label{UltimaComModulos}
\end{equation}
We can split the integrand in Eq.(\ref{UltimaComModulos})
as a composition of kernels in the regions where $k\gg p$ and $k\ll p$, yielding
\begin{equation}
\begin{split}
    \Sigma(p)=\frac{\alpha}{\pi p}\int_0^{\Lambda} dk \frac{k\Sigma(k)}{k^2+\Sigma^2(k)}\\\Big[{\cal K}_{k\gg p}(k,p) \Theta (k-p)+\\+ {\cal K}_{k\ll p}(k,p) \Theta(p-k)\Big],
    \end{split}
\end{equation}
where $\Theta(x)$ is the Heaviside function and ${\cal K}_{k\gg p}(k,p)$ and ${\cal K}_{k\ll p}(k,p)$ are  given by

\begin{equation}
  {\cal K}_{k\gg p}(k,p)\approx 2p-\frac{2p\theta^2}{k^2+\theta^2}
  \end{equation}
and
\begin{equation}
  {\cal K}_{k\ll p}(k,p)\approx 2k-\frac{2k\theta^2}{p^2+\theta^2}.
  \end{equation}
As a consequence, the integrals are separated into 
\begin{equation}\label{Sigma1}
\begin{split}
\Sigma (p) = \frac{\alpha}{\pi p}  \int_0^p   \frac{2k^2 \Sigma (k) }{k^2+\Sigma^2 (k)} \Big[1-\frac{ \theta^2}{k^2+\theta^2} \Big]   dk+\\ +\frac{\alpha}{\pi p}   \int_p^\Lambda   \frac{2k p \Sigma (k) }{k^2+\Sigma^2 (k)} \Big[ 1-\frac{ \theta^2}{k^2+\theta^2}\Big] dk.
\end{split}
\end{equation}

 After taking the derivative of Eq.~(\ref{Sigma1}) with  respect to $p$ and
using the Leibniz integral rule \cite{CSBPQED,nascimento2015chiral,PhysRevD.96.034005}, it becomes a nonlinear differential equation given by
\begin{equation} \label{EDOsemAprox}
p^2\Sigma'' (p)+2p\Sigma' (p) + \frac{ 2 \alpha}{\pi }    \frac{p^2-\frac{ p^2\theta^2}{p^2+\theta^2}   }{p^2+\Sigma^2 (p)}    \Sigma (p) =0.
 \end{equation}
This is supplemented by two boundary conditions, namely,
\begin{equation}
    \lim_{p\to\Lambda} \left[
    p \frac{d\Sigma(p)}{dp}+\Sigma(p)
    \right]=0\label{UV}
\end{equation}
for the UV regime and
\begin{equation}
\lim_{p\to0}  p^2 \frac{d\Sigma(p)}{dp}=0\label{IR}
\end{equation}
for the infrared regime.

\begin{figure}[H]
    \centering
    \noindent\includegraphics[width=.43\textwidth]{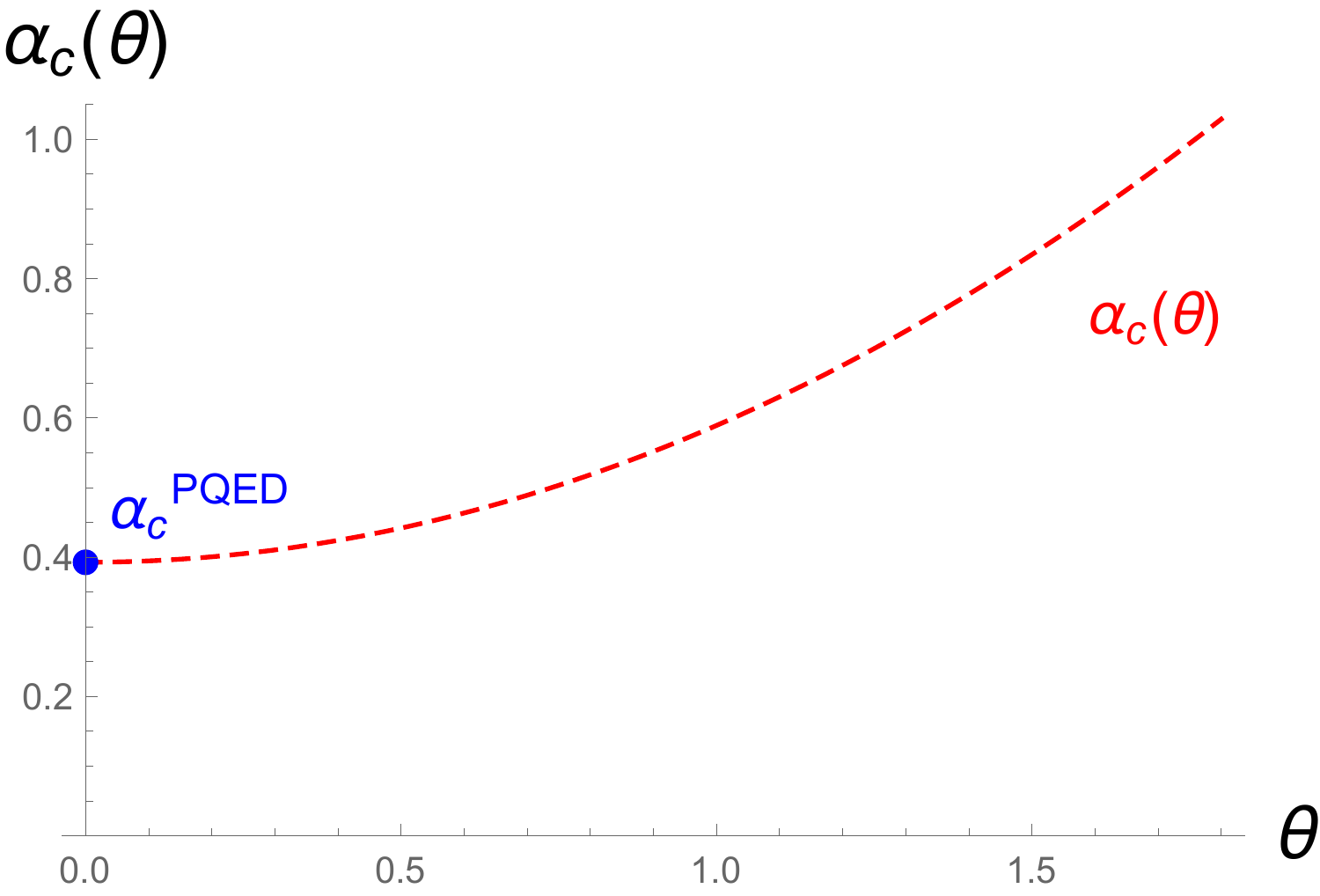}
  \caption{The critical fine-structure constant as a function of $\theta$. We plot Eq.~(\ref{alphac}) with $\Lambda=10$ (units of energy). The blue dot is the critical fine-structure  
constant of PQED at $\theta=0$, which has been discussed in 
Ref.~\cite{CSBPQED,nascimento2015chiral,PhysRevD.96.034005}. The dashed-red line shows that as we increase $\theta$,  the critical constant does also increase,  eventually bringing the system onto its symmetric phase where the mass function vanishes.}
    \label{alphaC}
\end{figure}

Equation (\ref{EDOsemAprox}) has a nonlinear behavior  in $p$, which prevent us of calculating an analytical solution for the mass function. This nonlinear feature is generated by the last term of the lhs of Eq.(\ref{EDOsemAprox}). However, we may find two linear differential equation by 
assuming either $p\ll \Sigma(p)$ or $p\gg \Sigma(p)$. For the first assumption, nevertheless, it is shown that the corresponding solution does not obeys the boundary conditions and, therefore, this is an unphysical solution. For the second case, we assume that $p\approx\Lambda$ to find
\begin{equation} 
p^2\Sigma'' (p)+2p\Sigma' (p) + \frac{ 2 \alpha}{\pi }    { \Big[1-\frac{ \theta^2}{\Lambda^2+\theta^2} \Big]  }    \Sigma (p) =0.\label{SDEaprox}
 \end{equation}
 This is a kind of Euler differential equation, whose solutions read 
  \begin{equation}\label{sig}
      \Sigma (p) = A p^{\lambda_+}+B p^{\lambda_-},
 \end{equation}
 with
 \begin{equation}
     \lambda_\pm=\frac{-1\pm\sqrt{1-4\left[
      \frac{ 2 \alpha}{\pi \varepsilon}   { \Big(1-\frac{ \theta^2}{\Lambda^2+\theta^2} \Big) }   
     \right]}}{2},
 \end{equation}
 where the critical coupling constant reads
 \begin{equation}
     \alpha_c(\theta)= \frac{\pi\varepsilon}{8\Big(1-\frac{ \theta^2}{\Lambda^2+\theta^2} \Big)}=\alpha^{PQED}_c\left(1+\frac{\theta^2}{\Lambda^2}\right),
     \label{alphac}
 \end{equation}
indicating that the criticality obtained for the base PQED can effectively be controlled by the CS mass parameter $\theta$, as seen in Fig.~\ref{alphaC}. 

The fact that $\Lambda$ does also appear in Eq.~(\ref{alphac}) is a consequence of the scale invariance of the model when $\theta$ vanishes.  In particular, when we take the continuum limit ($\Lambda\rightarrow\infty$), the role of $\theta$ also disappears, as expected.
Indeed, this represents a continuum limit, where the dynamical mass generation occurs for any value
of $\alpha$ and, therefore, we do not have an actual phase transition ~\cite{CSBPQED,nascimento2015chiral,PhysRevD.96.034005}.

\begin{figure}[t]
    \centering
    \noindent\includegraphics[width=.43\textwidth]{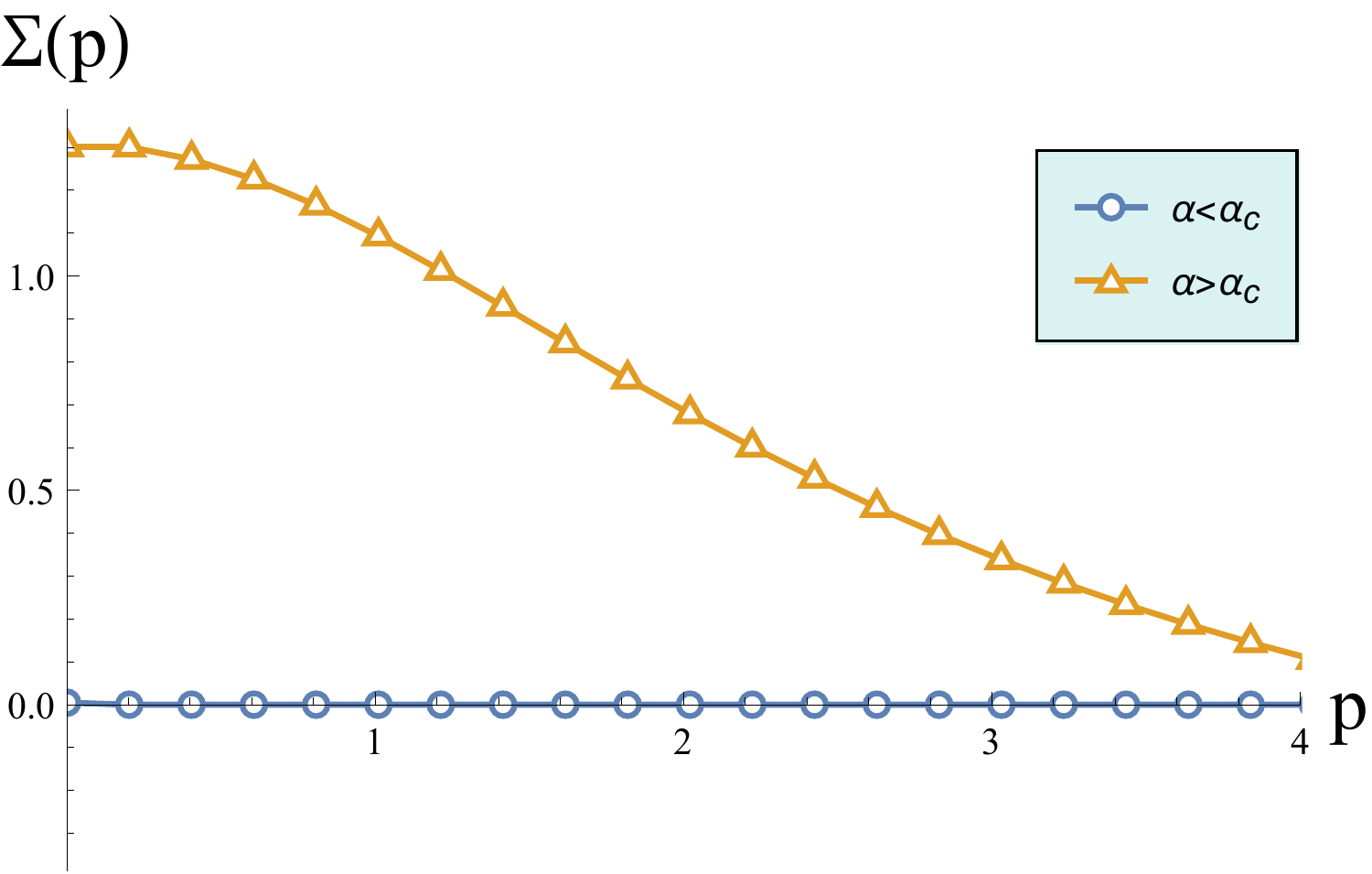}
  \caption{The mass function generated by PCS model above and below its critical point $\alpha_c(\theta)$. Plotted numerical solution of Eq.~(\ref{UltimaComModulos}) with  $\theta=0.1$, $\Lambda=10$, and hence $\alpha_c\approx.39$.  The blue line ($\alpha=0.2<\alpha_c(\theta)$) shows that dynamical mass generation is negligible  below  the critical fine-structure parameter,  while the yellow line exhibits the typical behavior above its critical value ($\alpha=0.8>\alpha_c$).}
    \label{alpha01}
\end{figure}

\begin{figure}[h!]
    \centering
    \noindent\includegraphics[width=.43\textwidth]{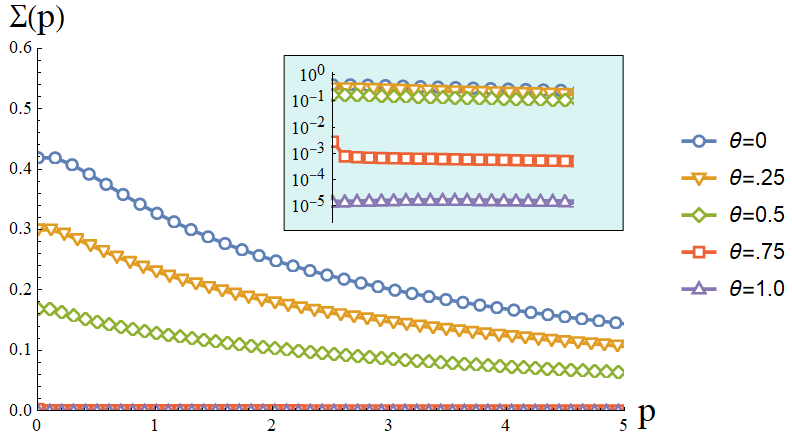}
    \caption{The mass function generated by PMCS model for different values of the PCS parameter $\theta$. Plotted  numerical solution of Eq.~(\ref{UltimaComModulos}) assuming $\alpha=0.8>\alpha_c(\theta)$ and $\Lambda=10$. The blue inset is the log scale of the same plot to illustrate that the mass generated gets even smaller as $\theta$ approaches 1.}
    \label{Vartheta}
\end{figure}

Even though $\Sigma(p)$ analytical solutions could only be obtained approximately
, the integral Eq.~(\ref{UltimaComModulos}) needs no approximation to be numerically evaluated. Hence, we can apply the same numeric calculation described in Refs.~\cite{CSBPQED,luis} to get a glimpse of the features related with the mass generation at the PMCS model, as illustrated in Figs.~\ref{alpha01}~and~\ref{Vartheta}.

The critical behaviour around $ \alpha_c$ can be contrasted by establishing a straightforward comparison between the blue and the yellow lines in Fig.~\ref{alpha01}. Below the critical value $\alpha_c$, the magnitude of the mass generated is significantly ($10^6$ times) weaker  than the mass generated above it.  For $\Lambda=10 eV$, 
 we conclude that the  mass generated  below $\alpha_c$ would be in the order of $10^{-7}\Lambda$, which is negligible for graphene, since $\Lambda_{g}\approx 3\; eV$ (see \cite{grapexp,novoselov2005two,geim2010rise,review}).
 

In this section we have neglected the quantum corrections to the gauge-field propagator. Nevertheless,
we may go beyond this by introducing the large-N expansion, which is sometimes called unquenched-rainbow approximation. Next, we consider this case.
\section{Unquenched PMCS} \label{unquenchedSec}

PQED has been shown to have a critical number of fermions $N_c$ which separates the symmetric from the broken phase \cite{CSBPQED,nascimento2015chiral,PhysRevD.96.034005}, entailed when the number of copies of the fermionic field $N$ is greater  than $N_c$. In this section, we  investigate the role of the CS term in the criticality of PQED in the large-$N$ regime. Within the unquenched-rainbow approximation, we must extend the fermionic sector in Lagrangian (\ref{LMCS}) to $N$ copies of massless fermions field,thus, we  consider 

\begin{equation}
    \begin{split}
        {\cal L}_{\mathrm{PMCS}}=\frac{1}{2} \frac{F_{\mu\nu}F^{\mu\nu}}{\sqrt{-\Box}}
+\frac{i\theta }{2} \frac{\epsilon_{\mu\nu\gamma}A^\mu \partial^\nu A^\gamma}{\sqrt{-\Box}}+
\\+\sum_{a=1}^{N} \Bar{\psi}^a\Big(i \slashed{\partial}+e\gamma^{\mu}A_{\mu}\Big)\psi^a 
+\frac{\lambda}{2} \frac{(\partial_\mu A^\mu)^2}{\sqrt{-\Box}}.
    \end{split}\label{LMCS2}
\end{equation}

\begin{figure}[htbp]
    \centering
    \noindent\includegraphics[width=.35\textwidth]{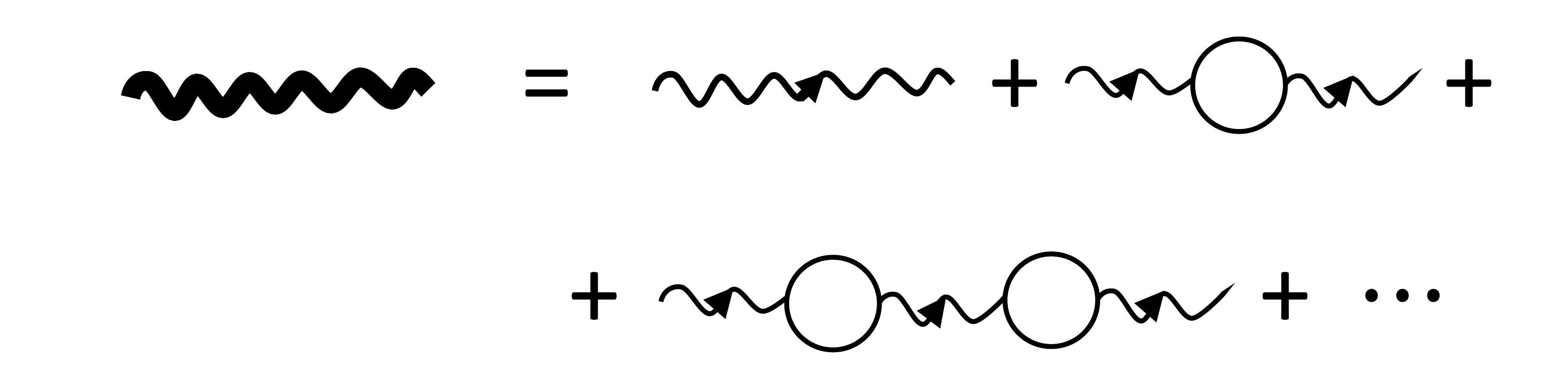}
  \caption{Dominant order of the $1/N$ expansion of a photon propagator.}
    \label{Bolhas}
\end{figure}

Applying a similar approach, it is possible to generalize such result to the PMCS model by adding up the dominant order of the $1/N$ expansion ($\equiv$ random phase approximation \cite{fernandez2021dynamical}, see Fig.~\ref{Bolhas})  to the truncated gauge-field propagator in Sec.~\ref{quenchedSec}, the gauge-field propagator for the PMCS model would then be
\begin{equation}
    \Delta_{\mu\nu}= T P_{\mu\nu}+ L\epsilon_{\mu\nu\rho} p^\rho\label{unquench}
\end{equation}
with 
\begin{eqnarray}
   T&=&\dfrac{8[ (p^2)^{3/2}+ ( \sqrt{p^2}-\tfrac{1}{2})\theta^2]
   }{[4(p^2-\theta^2)+\Pi_1^2]^2+64p^2\theta^2}
   +\nonumber\\&&
   +\dfrac{ \Pi_1[5p^2-(\Pi_1+\sqrt{p^2})^2]} {[4(p^2-\theta^2)+\Pi_1^2]^2+64p^2\theta^2},\nonumber\\
    L&=&\frac{2\theta}
    {4\left[(p^2)^{3/2}-p^2\Pi_1 +\theta^2\sqrt{p^2}\right]+\sqrt{p^2}\Pi_1^2}
    \end{eqnarray}
and
\begin{equation}
\Pi_1(p)=-\frac{e^2}{8}  N \sqrt{p^2}
\end{equation}
for massless fermions.         

Therefore, the gauge-field propagator in Eq.~(\ref{unquench}) simplifies to the PMCS propagator in Ref.~\cite{GabrielPCS}
\begin{equation}
    \Delta_{\mu\nu}(p)=\frac{p^2P^{\mu\nu}+\frac{16}{16+Ne^2}\theta\epsilon_{\mu\nu\rho}p^{\rho}}{2\varepsilon \left(1+\frac{Ne^2}{16}\right)\sqrt{p^2}\left[p^2+\frac{\theta^2}{(1
    +\frac{Ne^2}{16})^2}\right]}.
\end{equation}
 In order to compatibilize our model with the $1/N$ expansion, we apply the transformation $e\to e/\sqrt{N}$ to get
 \begin{equation}
       \Delta_{\mu\nu}=\frac{p^2P_{\mu\nu}+ \Tilde{\theta} \epsilon_{\mu\nu\rho}p^\rho}{2\varepsilon_{ef}\sqrt{p^2} (p^2+\Tilde{\theta}^2)},
 \end{equation}
 whereas the effective dielectric constant and effective CS parameter are, respectively,
 \begin{equation}
     \begin{split}
  \varepsilon_{ef}=& \,\varepsilon \left(1+\frac{e^2}{16}\right),
\\
     \Tilde{\theta}=&\,\frac{\theta}{1+\frac{e^2}{16}}.
    \end{split} \label{Effective}
\end{equation}
 
 Thus, similar to Eq.(\ref{SigmaSemQuench}), the mass function in the unquenched regime reads
 \begin{equation}
\Sigma(p)=\frac{e^2}{N} \int\frac{d^3k}{(2\pi)^3}\frac{\Sigma(k)\delta^{\mu\nu}\Delta_{\mu\nu}(p-k)}{k^2+\Sigma^2(k)},
 \end{equation}
 where we use $A(p)=1$ because $A(p)\approx 1+{\cal O}(1/N)$,  and we are calculating $\Sigma(p)$  only up to the dominant order of the $1/N$ expansion.

\begin{figure}[H]
    \centering
    \noindent\includegraphics[width=.48\textwidth]{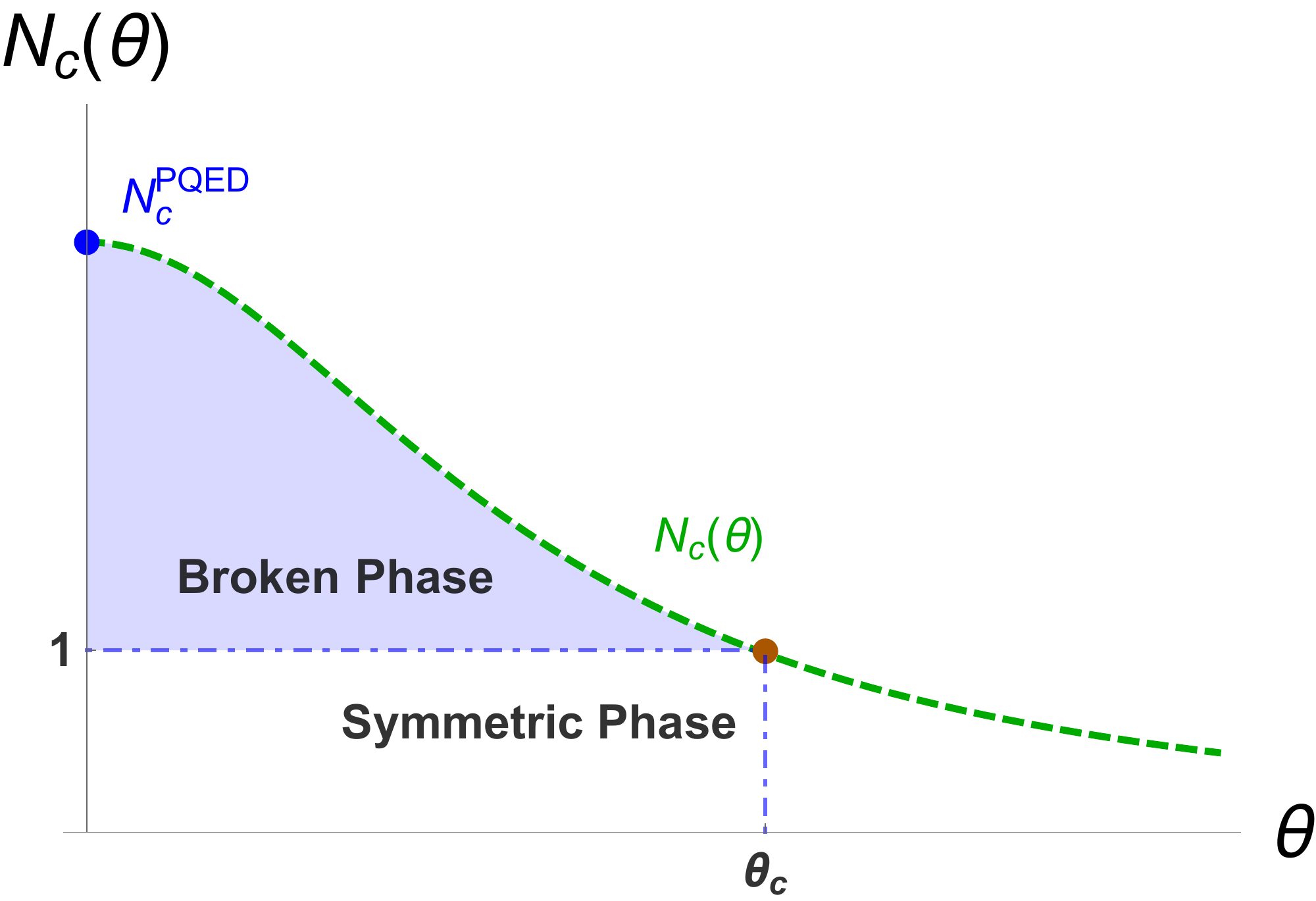}
  \caption{Sketch of the maximum number of flavors that allow a manifestation of a broken phase as a function of $\theta$.   The blue dot represents the critical number of flavors of PQED, where $\theta=0$ deduced in 
Ref.~\cite{CSBPQED,nascimento2015chiral,PhysRevD.96.034005} for $\varepsilon=1$. The dashed-green line shows that as we increase $\theta$,  the critical number of fermions  decreases, guiding the system onto its symmetric phase when $N_c$ gets smaller than one.}
    \label{NCsket}
\end{figure}

 Since the propagator in Eq.~(\ref{Photonprop}) has the same overall structure as the PMCS propagator in Ref.~\cite{GabrielPCS}, the only associations necessary to map the quenched approximation in Sec.~\ref{quenchedSec} to an unquenched PMCS model are
  \begin{equation}
     \begin{split}
     \frac{2\alpha}{\pi\varepsilon}\to&\ \frac{e^2}{2\pi^2\varepsilon_{ef}N},\\
     \theta \to&\  \Tilde{\theta}.
    \end{split}\label{map}
\end{equation} 
 Applied to Eq.~(\ref{SDEaprox}), the transfomations~(\ref{map})  lead to
\begin{equation} 
p^2\Sigma'' (p)+2p\Sigma' (p) +\frac{e^2}{2\pi^2\varepsilon_{ef}N}   { \Big(1-\frac{ \Tilde{\theta}^2}{\Lambda^2+\Tilde{\theta}^2} \Big)  }    \Sigma (p) =0,
 \end{equation} 
which satisfies the same boundary conditions as Eqs.~(\ref{UV}, \ref{IR}). This allows us to identify a chiral symmetry broken phase (shaded region in Fig.~\ref{NCsket}) up to a critical number of fermions   
 \begin{equation}
     N_c(\theta)=\frac{2e^2}{\pi^2\varepsilon_{ef}}   \left(1-\frac{\Tilde{\theta}^2}{\Lambda^2+\Tilde{\theta}^2}\right)=N_{c}^{PQED}\left(\frac{\Lambda^2}{\Lambda^2+\Tilde{\theta}^2}\right) \label{NC}
 \end{equation}
   directly affected by the PCS parameter $\theta$. Eq.~(\ref{NC}) does also
   exhibit the same criticality as PQED when we
turn oﬀ the $\theta$ (blue point in Fig.~\ref{NCsket}) and in the
strong-coupling limit, where they also reproduce QED3 for $\theta=0$.

Notice that $N_c$ must be at least larger than one in order to obtain a broken phase, otherwise the condition $N<N_c$ may not be satisfied. Such minimum value for $N_c$ to permit a symmetry-broken phase ($N_c\to1$) also limits the maximum value for the effective PCS mass parameter $\tilde\theta$ to 
\begin{equation}
 \Tilde  {\theta}<\Tilde{\theta}_c=\Lambda\sqrt{N_{c}^{PQED}-1},    \label{ThetaTilC}
\end{equation}
clearly setting this situation apart from the dynamics described when $\tilde\theta\geq\tilde\theta_c$, in which the system forcibly falls upon its symmetric phase, as we see in Fig.~\ref{NCsket}.

\section{Final remarks and perspectives}
\label{Resultados}

In this work, we have described the chiral symmetry breaking in the nonperturbative regime of the PMCS electrodynamics, a renormalizable model composed by associating PQED to a massive PCS parameter $\theta$~\cite{GabrielPCS,PLBOzela}. We have written the  Schwinger-Dyson equations for the model assuming the planar Dirac fermions were 4-component spinors  and  have applied the rainbow-quenched and the  unquenchend approximations to determine the criticality within the system. 

At the rainbow-quenched approximation, we have deduced an analytical expression for the dynamical mass  $\Sigma(p)$ [Eq.~(\ref{sig})] and for the critical fine-structure constant $\alpha_{c}$ [Eq.~(\ref{alphac})], emphasizing their dependence on $\theta$ 
to exhibit a correction (proportional to $\theta^2/\Lambda^2$) the PCS term brings to the hitherto known $\alpha_{c}^{PQED}$ -- whence we deduce that
 $\alpha_ c(\theta,\Lambda)$  grows as we enhance the PCS contribution even though such criticality correction diminishes conversely at larger energy scales, effectively retrieving $\alpha_{c}^{PQED}$. It is also worth to note that very large $\theta$ values inhibit the dynamical mass generation since this brings the system to ${\alpha}<{\alpha_c}$, favoring the symmetric phase. 

At the unquenched approximation,  we have considered  $N$ copies of the Dirac spinors to calculate the critical number of flavors $N_c$, governed by Eq.~(\ref{NC}). The dependence of  the critical parameter on the PCS term in this approximation works the other way around -- \textit{i.e.}, $N_c$ decreases towards zero as $\theta$ grows for a given $\Lambda$ -- by setting an upper value for $N$ and $\theta$ (see Eq.~\ref{ThetaTilC}). 
Whilst, the vacuum polarization  onto the gauge field propagator not only produces a screening effect, but also modifies the mass of the gauge field according to Eq.~(\ref{Effective}). In the strong-coupling limit, this mass disappears  and then $\Sigma(p)\to\Sigma(p)^{PQED}$.
Comparing the $N_c$ results for the PMCS electrodynamics with the usual Maxwell-Chern-Simons model, see appendix, discloses that they both have the same  structure.
Differently from the PMCS, however,  the Maxwell-Chern Simons model has a characteristic energy scale (defined by its fine-structure constant), which reflects the fact that this model is super renormalizable. 

Both approximations reproduce PQED in the continuum
limit, since $\theta\ll\Lambda\to\infty$. These recursive results regarding PQED are also obtained once we turn the PCS mass off ($\theta\to0$),  retrieving $\alpha_{c}\to\alpha_{c}^{PQED}$ and $N_{c}\to N_{c}^{PQED}$. Thence, we hope that the PMCS model described here will inherit the utility of PQED at the characterization of different kinds of two-dimensional materials, but now with the extra feature of including superconductivity and bound states, as happens in magic-angle twisted bilayer graphene~\cite{song2019all,PhysRevB.104.L121405} and  hybrid graphene/LiNbO3 platforms~\cite{yu2020hybrid}.



 \section*{Acknowledgements}

R.F.O. and G.C.M are partially supported by Coordena\c{c}\~ao de
Aperfei\c{c}oamento de Pessoal de N\'{\i}vel Superior -- Brasil
(CAPES), finance code 001; R.F.O., V.S.A., and L.O.N. by CAPES/NUFFIC, finance code 0112; and
V.S.A. and L.O.N.  by research grants from Conselho
Nacional de Desenvolvimento Cient\'{\i}fico e Tecnol\'ogico (CNPq).


\appendix


\section{\textbf{The QED with CS term.}}
\label{appA}
In this appendix, we discuss the main results regarding the dynamical symmetry breaking of MCS theory~\cite{Kondo,hong1993dynamical,williams1994dyson}. Our main idea is to show that the $\bar\theta$-parameter also favors the symmetric phase. The action of the model, in the Euclidean space, reads
\begin{eqnarray}
{\cal L}_{\rm MCS}&=&\frac{1}{4}\bar F^{\mu\nu}\bar F_{\mu\nu}+\frac{\lambda}{2}(\partial^\mu \bar A_\mu)^2- i \frac{\bar\theta}{2} \epsilon_{\mu\nu\alpha}\bar A^\mu\partial^\nu \bar A^\alpha +\nonumber\\
&+&\bar\psi_a(i\gamma^\mu\partial_\mu+\bar e\gamma^\mu \bar A_\mu)\psi_a, \label{A1}
\end{eqnarray}
where $\bar A_\mu$ is a gauge field (usually called the Maxwell-Chern-Simons field), $\bar\theta$ is the CS parameter with unit of mass $[\bar\theta]=M$, $\bar e$ is the coupling constant with unit given by $[\bar e]=M^{1/2}$, $\lambda$ is the gauge-fixing parameter, and $\bar F_{\mu\nu} =\partial_\mu \bar A_\nu-\partial_\nu \bar A_\mu$ is the strengh field tensor of $\bar A_\mu$. The matter field is given by the Dirac field $\psi$ and $\gamma_\mu$ are the four-rank Dirac matrices, obeying the same properties as in Sec.~II-III. Here, we consider a flavor index $a=1,...,N$ which shows that we have $N$ copies of the matter field. 

We shall follow the same steps as we have done in the case of the PMCS model. Hence, let us us summarize the main results. Within the large-$N$ expansion, at lowest order, the gauge-field propagator reads
\begin{equation} 
\Delta_{\mu\nu}(p)=\Delta(p)\left(\delta_{\mu\nu}-\frac{p_\mu p_\nu}{p^2}\right)+L(p)\epsilon_{\mu\alpha\nu}p^\alpha, \label{A2}
\end{equation}
where
\begin{equation}
\Delta(p)=\frac{p^2+\frac{\bar e^2}{8} |p|}{p^2[(|p|+\frac{\bar e^2}{8})^2+\bar\theta^2]} \label{A3}
\end{equation}
and
\begin{equation}
L(p)=\frac{\bar\theta}{[(p^2)^{2}+2(p^2)^{3/2}\frac{\bar e^2}{8}+p^2\frac{\bar e^4}{64}+p^2\bar\theta^2]}. \label{A4}
\end{equation}
Using Eq.~(\ref{A3}) and Eq.~(\ref{A4}) in the Schwinger-Dyson equation of the electron propagator, one may conclude, after some work, that the mass function $\Sigma(p)$ obeys an integral equation, given by
\begin{eqnarray}
\Sigma(p)=\frac{\bar e^2}{4N\pi^2}\int_0^\infty dk \frac{k\Sigma(k)}{k^2+\Sigma^2(k)}\nonumber\\
\frac{1}{p}\ln\left[\frac{(p+k+\frac{\bar e^2}{8})^2+\bar\theta^2}{(|p-k|+\frac{\bar e^2}{8})^2+\bar\theta^2}\right].\label{A5}
\end{eqnarray}
Equation (\ref{A5}) admits an expansion in its kernel, such that one may convert this integral equation into a differential equation.
To achieve that, we expand the logarithm in Eq.~(\ref{A5}), assuming  $\bar e^2\gg p$ (because chiral symmetry breaking usually occurs for strong interactions), to derive the property
\begin{eqnarray}
\ln\left[\frac{\left(p+k+\dfrac{\bar e^2}{8}\right)^2+\bar\theta^2}{\left(\vert p-k\vert+\dfrac{\bar e^2}{8}\right)^2+\bar\theta^2}\right]\approx \frac{
32k\bar e^2 \, \Theta(p-k)}{64k^2+\bar e^4+64\bar\theta^2}+\nonumber\\
+
\frac{32p(8k+\bar e^2) \Theta(k-p)}{64k^2+\bar e^4+64\bar\theta^2}\hspace{2mm} \label{aproxlogqedcs}.
\end{eqnarray}
This approximation has also been applied for the usual QED3 in Refs.~\cite{pisarski1984chiral,appelquist1986spontaneous,appelquist1988critical,nash1989higher,hoshino1989dynamical,dagotto1989computer,dagotto1990chiral,pennington1991masses,curtis1992dynamical,atkinson1990dynamical,kondo1992cutoff,burden1991light} and in the QED3 coupled to the usual CS term in Refs.~\cite{hong1993dynamical,williams1994dyson,Kondo}. 

Applying the property in  Eq.~(\ref{aproxlogqedcs}) to the  MCS mass function in Eq.~(\ref{A5}), we get
\begin{eqnarray}
\Sigma(p)=\frac{\bar e^2}{4\pi^2N}\left[\int_0^p dk\frac{ k^2\Sigma(k)}{k^2+\Sigma^2(k)}\frac{1}{p}\frac{32\bar e^2}{64k^2+\bar e^4+64\bar\theta^2}+\right.\nonumber\\\left.
+\int_p^\infty dk \frac{ k\Sigma(k)}{k^2+\Sigma^2(k)}\frac{32(8k+\bar e^2)}{64k^2+\bar e^4+64\bar\theta^2}\right],\nonumber\\\label{eqint2}
\end{eqnarray}
which can, in turn, be translated to the second order differential equation, whichs linearized version for $p^2\gg\Sigma^2(p)$, is given by 
\begin{equation}
\frac{d}{dp}\left[p^2\frac{d\Sigma(p)}{dp}\right]+\frac{8\bar e^4}{N\pi^2}\frac{\Sigma(p)}{\bar e^4+64\bar\theta^2}=0 \label{eqdifqedcs}
\end{equation}
supplemented by two asymptotic conditions similar to Eq.~(\ref{UV}) and Eq.~(\ref{IR}).

With that in mind, we obtain the critical number of fermions $N_c$, given by 
\begin{equation}
N_c(z)=N_c^{QED3} \left(\dfrac{\bar e^4}{\bar e^4+64\bar\theta^2}\right)=N_c^{QED3}\frac{1}{(1+z^2)}, \label{ncqedcs}
\end{equation} 
where $z^2=64\bar\theta^2/\bar e^4$. Interesting to highlight that within the strong-coupling limit, \textit{i.e.}, $\bar e^2\gg \bar\theta$, we obtain $N_c^{QED3}\rightarrow 32/\pi^2\approx 3.24$ as in Eq.~(\ref{NC}). Thence, solving the equation $N_c(z_{c})=1$, we obtain $z_c=\sqrt{32/\pi^2-1}$, which in terms of $\bar\theta$ implies that we must have $\bar\theta<\bar\theta_c$, where $\bar\theta_c=z_c \bar e^2/8$. Therefore, we conclude that the presence of $\bar\theta\neq 0$ decreases $N_c$ and favors the symmetric phase.


\bibliography{refs}
\clearpage

\end{document}